\documentclass{article}

\usepackage{arxiv}

\usepackage{cite}
\usepackage{amsmath,amssymb,amsfonts}
\usepackage{graphicx}
\usepackage{url}
\usepackage{textcomp,nicefrac}
\usepackage{booktabs}
\usepackage{multirow}
\usepackage[hidelinks]{hyperref}
\usepackage[switch]{lineno}

\begin{document}

\title{Integer Quantization of Graph Neural Networks for Real-Time FPGA Track Finding}

\author{Andrea Cardini, Pelayo Leguina, Elena Aller Gutierrez, Santiago Folgueras \\[0.2cm]
Universidad de Oviedo, Spain \\ and \\ Instituto de Ciencias y Tecnol\'ogicas Espaciales de Asturias (ICTEA)
}

\maketitle

\begin{abstract}

Real-time track finding for displaced-muon signatures in the CMS Level-1 trigger must operate under strict fixed-latency constraints of 12.5 $\mu$s while processing high-throughput detector data.  Because muon hits map naturally onto sparse, irregular graphs, graph neural networks (GNNs) are attractive candidates; however, mapping message-passing models to field-programmable gate arrays (FPGAs) requires careful co-design of numerical precision, microarchitecture, and high-level synthesis (HLS) implementation.
This work presents a reproducible, bit-exact workflow bridging GNN design and FPGA prototyping for fixed-latency inference. The methodology is demonstrated by implementing a two-layer GraphSAGE network onto an XCVU13P FPGA, using the Cora citation network as a fixed-size benchmark for firmware evaluation that decouples deployment feasibility from the physics task. Starting from a 32-bit floating-point reference that exceeds the available FPGA resource budget, we derive an integer-only datapath through post-training quantization (INT8 weights and activations, INT32 biases), a power-of-two scale approximation that replaces rescaling multipliers with arithmetic shifts, and data-driven bit-width narrowing. Every stage is validated bit-exactly against a Python integer emulator in Vitis HLS C-simulation. The optimized INT8 power-of-two design achieves an inference latency of 19 clock cycles (52.8 ns at the nominal 360 MHz clock) at 20\% DSP, 6\% FF, and 27\% LUT utilization, with 75.0$\pm$1.1\% accuracy compared to the 78.0$\pm$0.8\% for FP32. Accuracy is reported as the average across multiple training seeds. The resulting workflow establishes a concrete, transferable path toward fixed-latency GNN-based track reconstruction in the CMS Level-1 trigger.
\end{abstract}


\section{Introduction}
\label{sec:introduction}

The High-Luminosity phase of the Large Hadron Collider (HL-LHC) at CERN~\cite{HLLHC} is planned to increase the instantaneous luminosity delivered to the experiments. This is achieved primarily through higher beam intensity and focusing, resulting in more proton-proton interactions per bunch crossing, referred to as pileup. At the Compact Muon Solenoid (CMS) experiment~\cite{CMS:2008xjf}, the average number of proton-proton interactions per bunch crossing, i.e., every 25 ns, is expected to increase from about 60 to 140--200. To select potentially interesting events while operating in this high-pileup environment, the hardware-level trigger system, also known as the Level-1 (L1) trigger, is undergoing its Phase-2 upgrade~\cite{CERN-LHCC-2020-004}. The upgraded on-detector and back-end electronics based on field-programmable gate arrays (FPGAs) will deliver higher-granularity detector primitives at greater bandwidth to trigger boards with substantially expanded on-chip compute resources, enabling the real-time execution of data-driven pattern-recognition algorithms within the fixed 12.5 $\mathrm{\mu s}$ L1 latency budget \cite{CERN-LHCC-2020-004}. Such algorithms could include machine-learning inference for the identification of compatible hits from muons traversing the detector, namely a track-finding algorithm.
Combining detector hits to reconstruct the trajectory of a muon naturally maps onto a graph, where detector hits and their geometrical compatibility are represented as nodes and edges, respectively. A graph neural network (GNN) was therefore chosen as the architecture for the displaced-muon reconstruction algorithm. For successful deployment, the GNN must operate within a 12.5 $\mu$s latency budget and fit within the available FPGA resource constraints. This manuscript presents a systematic evaluation of integer quantization, numerical representation, and high-level synthesis (HLS) implementation choices for a GNN-based inference within FPGA latency and resource constraints. The study is performed on a fixed-size firmware-proxy benchmark, decoupling deployment feasibility considerations from the physics application in the CMS L1 trigger.

\section{Firmware Requirements for the Phase-2 Level-1 Trigger}
\label{sec:firmware}

In the context of the HL-LHC, the L1 trigger system of the CMS experiment would have to process events at a rate of 40 MHz, producing an output rate of 750 kHz. Such events would then be sent to the software-level trigger for further processing and filtering. The maximum time available for the L1 decision is 12.5 $\mu$s, set by the on-detector buffer capacity before data are either discarded or transferred to the CPU farm.

This work focuses on the deployment of a track-finding algorithm in the muon system, and in particular in the overlap region between the barrel and endcap detectors. This region, extending approximately in the pseudorapidity range $0.8 < |\eta| < 1.24$, is characterized by the presence of three distinct detector technologies~\cite{MUOTDR}: drift tube (DT) chambers, cathode strip chambers (CSCs), and resistive plate chambers (RPCs). Currently, muon tracks in this region are reconstructed by the Overlap Muon Track Finder (OMTF)
algorithm~\cite{CMS:2019omtf,Zabolotny:2016omtf}, which combines information provided by the different detectors, such as position and bending direction. 

For the Phase-2 upgrade, the OMTF will run on a custom ATCA (Advanced Telecommunications Computing Architecture) board housing Xilinx Virtex UltraScale+ FPGA devices with high-speed optical transceivers. The study presented in this manuscript targets an XCVU13P FPGA from the Xilinx Virtex UltraScale+ family, which provides 12,288 digital signal processing (DSP) slices, 1,728,000 look-up tables (LUTs), 3,456,000 flip-flops (FFs), 360 Mb of on-chip UltraRAM memory, and 128 GTY transceivers capable of operating at up to 32.75 Gb/s. This device is a larger member of the same FPGA family than the XCVU9P device quoted in Ref.~\cite{CERN-LHCC-2020-004} as the design choice for the OMTF, which provides 6,840 DSP slices, 1,182,000 LUTs, and 2,364,000 FFs.

\section{Graph Network Architecture}
\label{sec:architecture}

Initial studies~\cite{Leguina:2025suh,Cardini:2025IML} suggested that architectures based on graph sample-and-aggregate operations (GraphSAGE)~\cite{Hamilton:2017inductive} or more general message passing neural networks (MPNNs) could provide good performance for the target use case, consistent with prior demonstrations of GNN inference within sub-microsecond FPGA latency budgets for real-time particle reconstruction~\cite{Iiyama:2020wap}. 
The GNN deployed in this study is a deliberately simplified two-layer GraphSAGE, chosen to establish a firmware deployment methodology, rather than to optimize physics performance.

\begin{figure}[hbt!]
    \centering
    \includegraphics[width=0.4\linewidth]{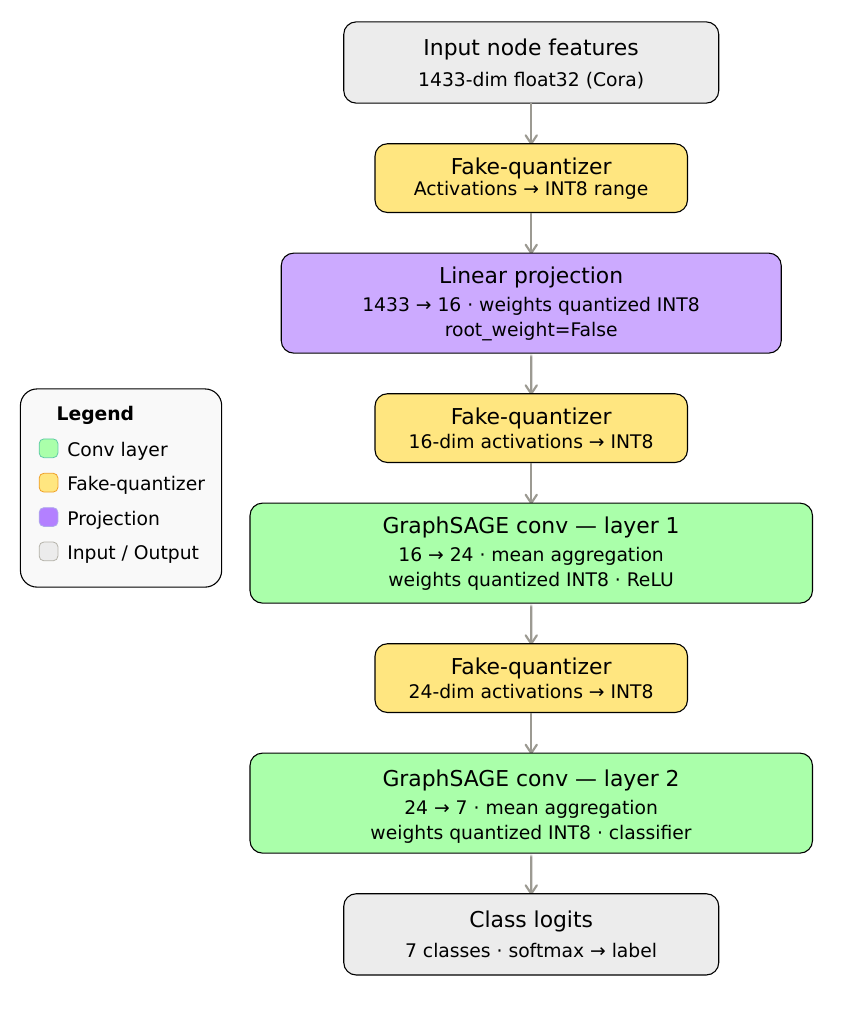}
    \caption{Schematic representation of the tested GraphSAGE-based network.}
    \label{fig:SAGEGNN}
\end{figure}

Fig.~\ref{fig:SAGEGNN} shows a schematic representation of the network constructed for firmware deployment studies: a two-layer GraphSAGE model~\cite{Hamilton:2017inductive} with mean neighbour aggregation, a linear input-feature projection, and statically bounded loops that permit full unrolling and deterministic pipelining in HLS~\cite{Duarte:2018ite,Iiyama:2020wap}.

Relative to the original GraphSAGE formulation~\cite{Hamilton:2017inductive}, the root-node self-term is omitted: each node embedding is updated solely from its aggregated neighbourhood, discarding the concatenation with the root representation. This choice eliminates a separate root weight matrix from the HLS datapath and reduces the number of multiply-accumulate (MAC) operations per node per layer, simplifying the fixed-size loop structure required for complete array unrolling~\cite{Duarte:2018ite}. The self-contribution can be reinstated as an additional integer accumulation at the output of the aggregation stage, making it a straightforward extension when deploying on the physics dataset.

\begin{figure}[hbt]
    \centering
    \includegraphics[width=0.6\linewidth]{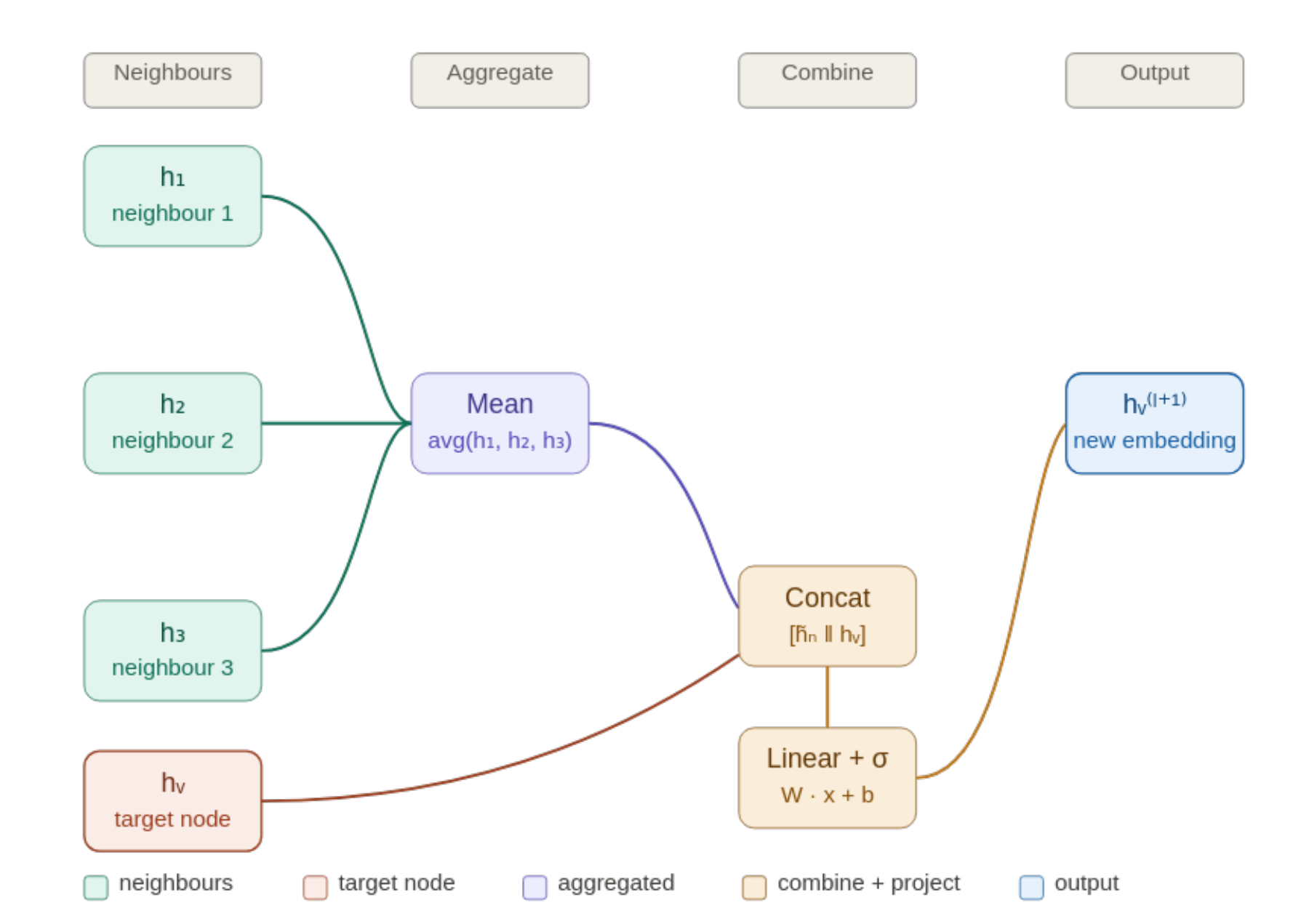}
    \caption{Schematic representation of one GraphSAGE layer.}
    \label{fig:SAGEsimple}
\end{figure}

Fig.~\ref{fig:SAGEsimple} shows a schematic representation of a GraphSAGE layer, where each node embedding is updated by aggregating information from its local neighbors.

\begin{figure}[h!]
    \centering
    \includegraphics[width=0.7\linewidth]{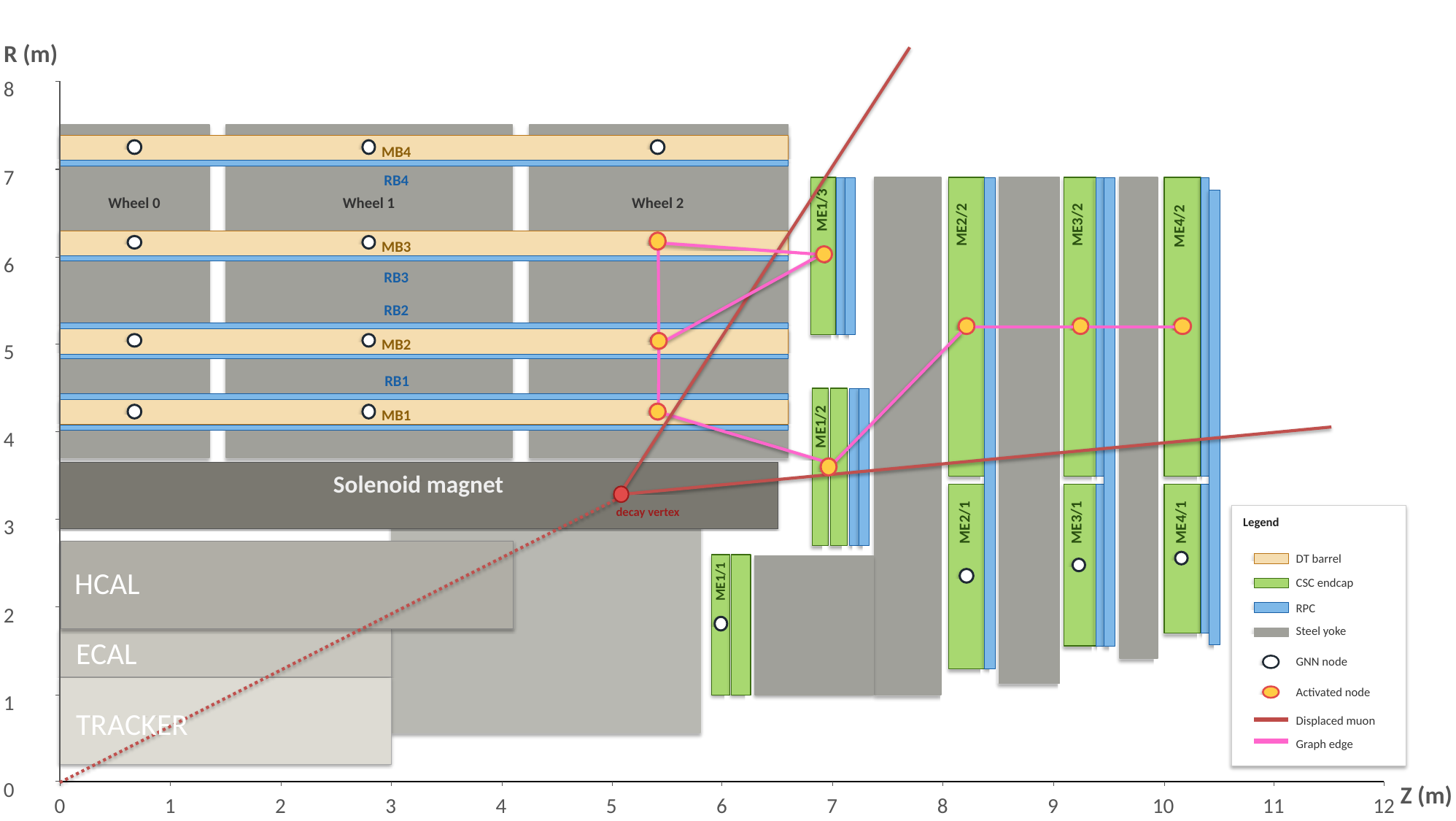}
    \caption{Schematic representation of the CMS muon system being traversed by a long-lived particle decaying into a pair of muons. The red lines represent the displaced muons, while the yellow circles represent activated detector hits that form a graph for the track-finding GNN.}
    \label{fig:examplegraph}
\end{figure}

The input data for the target GNN are represented as graphs. Detector stubs for each detection layer (DT, CSC, RPC)~\cite{CMS:2008xjf} are assigned to nodes using their coordinates, including radius $r$, pseudorapidity $\eta$, and azimuthal angle $\varphi$, together with the bending angle of the stub when it is recorded by a DT chamber. A maximum number of 8 detector layers (nodes) is considered per graph. Edges are constructed between stubs in detector regions that are geometrically compatible, meaning that a particle trajectory could plausibly cross both detector elements. A graphical representation of geometrically compatible hits forming a graph is presented in Fig.~\ref{fig:examplegraph}. 

The objective of the present work is not to evaluate the physics performance but to establish a methodology for firmware deployment. Therefore, the resource and latency requirements for the deployment of this graph network are evaluated on a benchmark dataset, as described in the next Section.

\section{Dataset and Data Preparation}
\label{sec:dataset}

The Cora citation network~\cite{Sen:2008collective} is used in this work as a firmware-proxy benchmark to develop and validate the quantization and FPGA deployment methodology. Cora consists of 2,708 scientific publications represented as nodes and 10,556 citation edges, with an average node degree of approximately 3.9. Each node is represented by a 1,433-dimensional binary bag-of-words feature vector and is labelled with one of seven research-topic classes. The standard Planetoid data splits~\cite{Yang:2016wmc} are adopted: 20 nodes per class, corresponding to 140 nodes in total, for training; 500 nodes for validation; and 1,000 nodes for testing. Before training, node features are row-normalized so that each feature vector sums to unity.

The full 2,708-node graph is not suitable for the fixed-size, fully unrolled HLS architecture considered in this deployment study. A compact fixed-size subgraph is therefore extracted for hardware-level validation: a 2-hop neighborhood around a representative training node is computed and truncated to $N_{\mathrm{dep}}$ nodes. The synthesised design uses $N_{\mathrm{dep}}=8$, matching the fixed eight-node graphs of the target muon application (Section~\ref{sec:architecture}). The adjacency matrix of the extracted subgraph is row-normalized without self-loops,
\begin{equation}
    \tilde{A}_{ij} = \frac{1}{\deg(i)},\quad (i,j)\in\mathcal{E},
    \label{eq:adj_norm}
\end{equation}
where $\deg(i)=\sum_{j:(i,j)\in\mathcal{E}}1$ is the number of incoming neighbours of node $i$, isolated nodes ($\deg(i)=0$) take $\tilde{A}_{ij}=0$, and $\mathcal{E}$ denotes the set of directed edges in the extracted subgraph, consistent with the mean-aggregation operation used in the deployed GraphSAGE layer.

The 1,433-dimensional Cora features would require approximately $1433\times F_h$ MAC operations per node per layer, with $F_h$ being the number of hidden features per node per layer. This is prohibitive for the fixed-latency inference architecture targeted here. A trainable linear projection layer reduces the input dimension to $F_{\mathrm{proj}}$ before the first graph convolution. This is precomputed off-chip and omitted from the model under study; its latency is therefore not included in the reported figures and must be absorbed upstream within the same fixed-latency pipeline. The FPGA-targeted model therefore operates with architecture $16{\to}24{\to}7$, corresponding to projected input features ($F_{\mathrm{proj}}=16$), hidden features, and output classes, respectively. The resulting model contains approximately $1.1\,\mathrm{k}$ weights, making it suitable for complete array unrolling in HLS.

\section{Quantization Methods}
\label{sec:methods}

Deploying floating-point GNNs on FPGAs is resource-prohibitive for the fully unrolled architecture considered in this study. As shown in Table~\ref{tab:quantization}, a 32-bit floating-point (FP32) implementation of the $16{\to}24{\to}7$ model reaches 283\% of the available XCVU13P DSP budget, 123\% of the available LUT budget, and 128\% of the available FF budget. A sequence of progressively optimized quantization schemes is therefore developed, each trading a bounded accuracy loss for substantial reductions in latency and resource usage.

\subsection{Float32 and Fixed-Point Baselines}
\label{sec:float_fixed}

A floating-point reference is obtained by converting the trained 64-bit PyTorch model to 32-bit floating point and synthesizing it with Vitis HLS using native \texttt{float} types. The resulting implementation is not deployable on a single XCVU13P device because the required DSP, LUT, and FF resources exceed the available budget.

A first quantized variant is obtained with the Xilinx \texttt{ap\_fixed<8,8>} arithmetic for activations and weights while retaining floating-point scale factors during dequantization. This is denoted with \textit{Fixed-PTQ} for ``fixed point - post-training quantization'' (PTQ). 

\subsection{Integer-Only Post-Training Quantization}
\label{sec:int8ptq}

A fully integer arithmetic pipeline is obtained by replacing all floating-point operations with integer MACs and fixed-point rescaling~\cite{Jacob:2018quant}. The trained floating-point model is calibrated on a representative subgraph to derive per-tensor symmetric quantization scales:
\begin{equation}
    s = \frac{\max|x|}{127},\quad
    q = \operatorname{clip}\!\left(\!\left\lfloor\frac{x}{s}\right\rceil,-128,\,127\right),
    \label{eq:int8_quant}
\end{equation}
where $\lfloor\cdot\rceil$ denotes rounding to the nearest integer, $x$ represents the floating-point tensor (weights or activations) being quantized, and $q$ the resulting 8-bit integer (INT8) weight matrices and INT8 activations. 

\textit{Calibration protocol:} The quantization scales are derived from a single forward pass over a fixed representative subgraph comprising the full 2-hop neighbourhood (a calibration subgraph of $N_{\mathrm{cal}}=32$ nodes) extracted around a training node. The scales are computed using symmetric per-tensor min/max observers, with the quantization grid aligned to zero (zero-point $= 0$). The calibration procedure evaluates all nodes in the extracted subgraph without resampling; as the subgraph is deterministic, derived scales are reproducible across runs. Quantization is applied symmetrically per tensor (not per channel), meaning a single scale applies to all elements of a weight matrix or activation map.

The row-normalized adjacency matrix is scaled to fixed-point integers as
\begin{equation}
    A^K_{ij} = \left\lfloor \tilde{A}_{ij}\cdot K \right\rceil,\quad K = 2^{K_b},\quad K_b = 12,
    \label{eq:adj_int}
\end{equation}
enabling full integer aggregation. Biases ($b_o$) are converted to the accumulator domain:
\begin{equation}
    b^{\mathrm{int}}_o = \left\lfloor \frac{b_o}{s_{\mathrm{hid}}\cdot s_w} \right\rceil,
    \label{eq:bias_int}
\end{equation}
where $s_{\mathrm{hid}}$ is the aggregation-output (hidden) activation scale consumed by the linear layer and $s_w$ is the weight scale, so that $b^{\mathrm{int}}_o$ lies in the same domain as the integer products $\hat{h}\,w$ accumulated in \eqref{eq:int_lin}.

The complete integer forward pass for each layer $l\in\{1,2\}$ of the deployed GraphSAGE model is:
\begin{align}
    T_{i,f}^{(l)} &= \sum_{j} A^K_{ij}\cdot h_{j,f}^{(l-1)}, \label{eq:int_agg}\\
    \hat{h}_{i,f}^{(l)} &= \operatorname{sat}_{8}\!\left(
    \left\lfloor \frac{T_{i,f}^{(l)}\cdot\beta^{(l)}_{\mathrm{fp}} + 2^{M-1}}{2^M} \right\rfloor
    \right), \label{eq:int_rescale_agg} \\
    a_{n,o}^{(l)} &= b_o^{(l)} + \sum_{f} \hat{h}_{n,f}^{(l)}\cdot w^{(l)}_{o,f}, \label{eq:int_lin}\\
    h_{n,o}^{(l)} &= \operatorname{sat}_{8}\!\left(
    \rho_l\!\left[
    \left\lfloor \frac{a_{n,o}^{(l)}\cdot\gamma^{(l)}_{\mathrm{fp}} + 2^{M-1}}{2^M} \right\rfloor
    \right]\right), \label{eq:int_rescale_lin}
\end{align}
The integer aggregation accumulator $T^{(l)}_{i,f}$ (node $i$, feature channel $f$, layer $l$) accumulates
the INT8 features $h^{(l-1)}_{j,f}$ of all its graph-neighbours $j$, weighted by
the fixed-point adjacency entries $A^K_{ij}$.  Because $A^K_{ij}$ are integers
scaled by $K = 2^{12}$, the partial sums $T^{(l)}_{i,f}$ span a range up to
$d_{\max}\cdot K\cdot 127$ and must be held in a wider accumulator before
rescaling.

Equation~\eqref{eq:int_rescale_agg} requantises $T$ to the INT8 range, yielding the
aggregated feature $\hat{h}^{(l)}_{i,f}$.
The product $T\cdot\beta^{(l)}_{\mathrm{fp}}$ is a fixed-point multiply: it
simultaneously undoes the $K$-scaling of the adjacency and converts between the
input quantization scale $s^{(l)}_{\mathrm{in}}$ and the aggregation-output scale
$s^{(l)}_{\mathrm{hid}}$.  Adding $2^{M-1}$ before the right-shift by $M$
implements round-to-nearest, while the outer $\mathrm{sat}_{8}$ clamps the
result to the signed INT8 range $[-128, 127]$.  With $M=24$ fractional bits, the largest scale multiplier (the layer-1 linear multiplier $\gamma^{(1)}_{\mathrm{fp}}\!\approx\!1.8\times10^{5}$) occupies 18 bits, fitting the 18-bit operand port of a DSP48E2 slice (a Xilinx DSP slice optimized for multiply-accumulate operations); the aggregation multipliers $\beta^{(l)}_{\mathrm{fp}}$ are narrower ($\le 13$ bits).

Equation~\eqref{eq:int_lin} is the standard matrix--vector step: the pre-scaled bias integer $b^{(l)}_o$ is added to the dot product of the rescaled aggregation output $\hat{h}^{(l)}$ and the quantized weight row $w^{(l)}_{o,\cdot}$, whose entry $w^{(l)}_{o,f}$ is the INT8 weight connecting input channel $f$ to output channel $o$. Both 
operands are INT8, so the accumulator $a^{(l)}_{n,o}$, where $n$ denotes the node index (equivalent to $i$ in the aggregation step), is INT32, large enough to hold up to $F = 24$ signed products without overflow, 

Equation~\eqref{eq:int_rescale_lin} brings the INT32 accumulator back to INT8 by
the same round-then-shift mechanism, now using $\gamma^{(l)}_{\mathrm{fp}}$.  The
layer activation function $\rho_l$ is a rectified linear unit (ReLU) for hidden layers and the identity for
the output layer; because ReLU only zeroes negative values, it can be applied
entirely in the integer domain before saturation, with no floating-point step.
\begin{align}
    \beta^{(l)}_{\mathrm{fp}} &= \left\lfloor \frac{s^{(l)}_{\mathrm{in}}}{K\cdot s^{(l)}_{\mathrm{hid}}} \cdot 2^M \right\rceil, \label{eq:beta_fp}\\
    \gamma^{(l)}_{\mathrm{fp}} &= \left\lfloor \frac{s^{(l)}_{\mathrm{hid}}\cdot s^{(l)}_w}{s^{(l)}_{\mathrm{out}}} \cdot 2^M \right\rceil. \label{eq:gamma_fp}
\end{align}
Here $s^{(l)}_{\mathrm{in}}$ denotes the scale of the activations entering the aggregation of layer $l$ (the projected-input scale for $l=1$ and the hidden scale $s_{\mathrm{hid}}$ for $l=2$), while $s^{(l)}_{\mathrm{hid}}$ and $s^{(l)}_{\mathrm{out}}$ are the aggregation-output and linear-output scales, respectively. Equations~\eqref{eq:int_rescale_agg} and \eqref{eq:int_rescale_lin} each require one wide-integer multiplication by a fixed-point scale, which maps to DSP48 slices in the synthesized design.

\subsection{INT8 with Power-of-Two Scaling}
\label{sec:po2}

The four scale multiplications in \eqref{eq:int_rescale_agg}--\eqref{eq:int_rescale_lin} consume a significant fraction of the available DSP budget. Constraining the scales to powers of two (PO2) eliminates these multiplications, replacing them with compile-time arithmetic right-shifts. In this work, each scale is approximated by its nearest power of two:
\begin{equation}
    \beta \;\approx\; \tilde{\beta} = 2^{-S_\beta},\quad
    S_\beta = \left\lfloor -\log_2\beta \right\rceil,
    \label{eq:po2_approx}
\end{equation}
Choosing the nearest power of two bounds the resulting multiplicative scaling error by a factor of $\sqrt{2}$. Under this constraint, \eqref{eq:int_rescale_agg} and \eqref{eq:int_rescale_lin} reduce to shift-and-round operations:
\begin{align}
    \hat{h}_{i,f}^{(l)} &= \operatorname{sat}_{8}\!\left(R_{S_{\beta}^{(l)}}\!\left(T_{i,f}^{(l)}\right)\right), \label{eq:po2_agg}\\
    h_{n,o}^{(l)} &= \operatorname{sat}_{8}\!\left(\rho_l\!\left[
    R_{S_{\gamma}^{(l)}}\!\left(a_{n,o}^{(l)}\right)
    \right]\right), \label{eq:po2_lin}
\end{align}
where $R_S(x)$ denotes round-to-nearest division by $2^S$, implemented on the signed accumulator as a single rounding addition followed by an arithmetic right-shift:
\begin{equation}
R_S(x)= \left\lfloor \frac{x + 2^{S-1}}{2^S} \right\rfloor = \left(x + 2^{S-1}\right) \gg S,
\label{eq:signed_shift_round}
\end{equation}
where $\gg$ is an arithmetic right-shift. This is the same rounding rule used in \eqref{eq:int_rescale_agg} and \eqref{eq:int_rescale_lin}, and rounds half-integers toward $+\infty$.
No multipliers are required for rescaling in this configuration.

For the two-layer $16{\to}24{\to}7$ model, the four shift amounts are derived analytically from the PTQ calibration scales:
\begin{align}
    S_{\beta^{(1)}} &= \left\lfloor-\log_2\!\left(\tfrac{s_{\mathrm{in}}}{K\cdot s_{\mathrm{hid}}}\right)\right\rceil, \label{eq:shift_beta1}\\
    S_{\beta^{(2)}} &= \left\lfloor-\log_2\!\left(\tfrac{1}{K}\right)\right\rceil = K_b = 12\quad(\text{exact}), \label{eq:shift_beta2}\\
    S_{\gamma^{(1)}} &= \left\lfloor-\log_2 s_{w_1}\right\rceil, \label{eq:shift_gamma1}\\
    S_{\gamma^{(2)}} &= \left\lfloor-\log_2\!\left(\tfrac{s_{\mathrm{hid}}\cdot s_{w_2}}{s_{\mathrm{out}}}\right)\right\rceil. \label{eq:shift_gamma2}
\end{align}
Notably, $S_{\beta^{(2)}}=K_b=12$ is exact because $\beta^{(2)} = s_{\mathrm{hid}}/(K\cdot s_{\mathrm{hid}}) = 1/K = 2^{-12}$ is already a power of two. The layer-1 linear scale reduces to $\gamma_1 = s_{w_1}$, as in \eqref{eq:shift_gamma1}, because the layer-1 aggregation output and the layer-1 linear output share the hidden scale ($s^{(1)}_{\mathrm{hid}} = s^{(1)}_{\mathrm{out}}$). For the reference model, the four shift constants are $S_{\beta^{(1)}}=17$, $S_{\beta^{(2)}}=12$, $S_{\gamma^{(1)}}=7$, and $S_{\gamma^{(2)}}=8$.
\begin{figure}[h!]
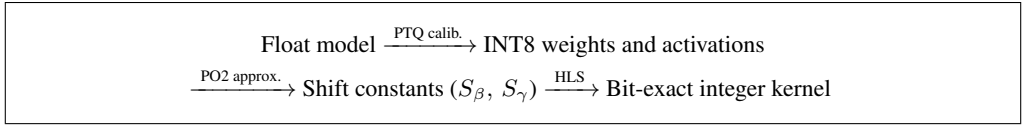

    \centering
    \fbox{\parbox{0.8\linewidth}{\centering\vspace{5pt}
        \small
        Float model
        $\xrightarrow{\;\text{PTQ calib.}\;}$
        INT8 weights and activations\\[3pt]
        $\xrightarrow{\;\text{PO2 approx.}\;}$
        Shift constants ($S_\beta,\,S_\gamma$)
        $\xrightarrow{\;\text{HLS}\;}$
        Bit-exact integer kernel
        \vspace{5pt}}}
    \caption{INT8-PO2 quantization and deployment pipeline. PTQ calibration on a representative subgraph yields per-tensor scales, which are approximated by powers of two. The resulting shift constants are embedded as compile-time macros in the Vitis HLS kernel, replacing the rescaling multiplications with arithmetic right shifts. Bit-exact agreement between the Python integer emulator and the HLS C-simulation is verified before synthesis.}
    \label{fig:po2_pipeline}
\end{figure}

A schematic representation of the INT8-PO2 pipeline is shown in Fig.~\ref{fig:po2_pipeline}. In the synthesized Vitis HLS kernel, shift amounts are encoded as preprocessor macros, such as \texttt{BETA1\_SHIFT}, \texttt{BETA2\_SHIFT}, \texttt{EFF\_SCALE1\_SHIFT}, and \texttt{EFF\_SCALE2\_SHIFT}, allowing the tool to resolve them into wiring and shift logic during elaboration.

\subsection{Data-Driven Bit-Width Optimization}
\label{sec:bwopt}

A 32-bit accumulator is sufficient to prevent overflow in any INT8 MAC chain considered here, but it wastes routing and flip-flop resources. A data-driven profiling pass runs an integer emulation of the full forward pass and records the maximum absolute value $m$ of each intermediate signal: linear accumulator $m_a$, aggregation temporary $m_T$, and adjacency entry $m_A$. The minimal sufficient bit-width for a signed integer type is then:
\begin{equation}
    B = \left\lceil\log_2(m+1)\right\rceil + 1 + \delta,
    \label{eq:bwidth}
\end{equation}
where $\delta=2$ is a safety margin and the extra $+1$ accounts for the sign bit.

The theoretical worst-case bounds,
\begin{equation}
    m_T^{\mathrm{th}} = d_{\max}\cdot A^K_{\max}\cdot 127,\quad
    m_a^{\mathrm{th}} = m_b + F\cdot 127\cdot w_{\max},
    \label{eq:bwidth_theory}
\end{equation}
where $d_{\max}$ is the maximum node degree in the subgraph, $A^K_{\max}\le K=2^{12}$ is the largest fixed-point adjacency entry, $w_{\max}$ and $m_b$ are the largest absolute INT8 weight and integer bias $|b^{\mathrm{int}}_o|$ respectively, $127$ is the maximum INT8 magnitude, and $F$ is the layer fan-in ($F^{(1)}=16$, $F^{(2)}=24$; Eq.~\eqref{eq:bwidth_theory} uses the worst case $F=24$). The adjacency magnitude requires no statistical bound: $A^K_{\max}\le K=2^{12}$
follows directly from \eqref{eq:adj_int}, so $m_A$ is fixed deterministically by
applying \eqref{eq:bwidth} with $m_A = A^K_{\max}$ rather than by profiling.

\begin{figure}[ht]
    \centering
    \includegraphics[width=0.8\linewidth]{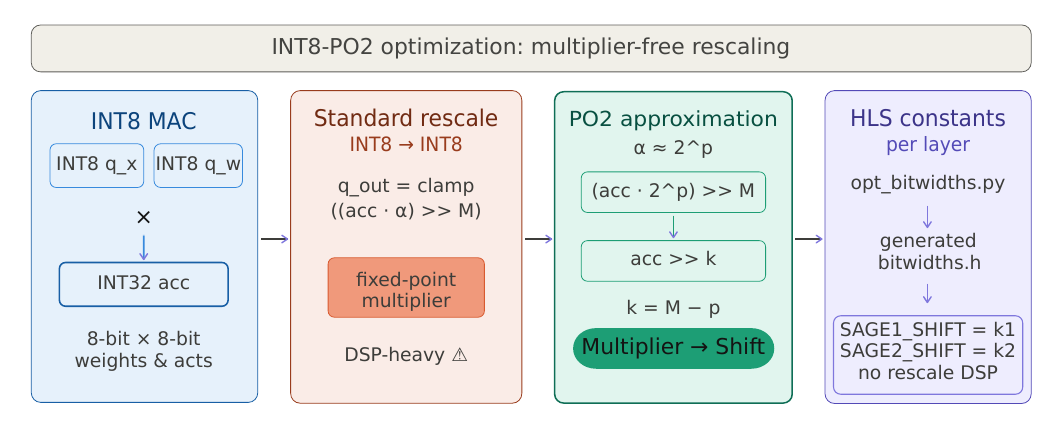}
    \caption{Schematic representation of the optimized integer quantization with power of two scaling: \textit{INT8-PO2 Opt.}}
    \label{fig:int8opt}
\end{figure}

Applied together with PO2 shift optimization, data-driven bit-width narrowing yields the \textit{INT8-PO2 Opt.} configuration, represented in Fig.~\ref{fig:int8opt}.

\subsection{Quantization-Aware Training}
\label{sec:qat}

Quantization-aware training (QAT) inserts differentiable fake-quantization operators around weight and activation tensors during the forward pass, allowing the model to adapt its parameters to the quantization grid under gradient descent. The Brevitas library~\cite{Pappalardo:brevitas} provides symmetric per-tensor INT8 fake quantizers with straight-through estimators and moving-average min/max observers. The QAT model is initialized from the pre-trained floating-point weights and fine-tuned for 200 epochs with quantization applied at INT8 precision to both weights and activations.

\section{Quantization and HLS Synthesis Results}
\label{sec:results}

The GraphSAGE model was tested using a custom graph-network implementation derived from the hls4ml package~\cite{Duarte:2018ite}. Tests were performed on the Cora dataset described in Section~\ref{sec:dataset}, using the quantization methods defined in Section~\ref{sec:methods}.

The model was evaluated under different quantization configurations in order to compare their impact on firmware-oriented performance. The metrics compared are:
\begin{itemize}
    \item latency, reported both in clock cycles and in nanoseconds at the nominal 360 MHz target clock frequency;
    \item FPGA resources required to deploy the model, namely DSP slices, FFs, and LUTs;
    \item estimated maximum clock frequency (Fmax) reported by HLS synthesis for an initiation interval (II) of 1;
    \item accuracy, estimated as the fraction of correctly classified nodes in the 1,000-node Cora test set.
\end{itemize}

All accuracy measurements are reported as the mean $\pm$ one standard deviation (std) over 5 random training seeds, with seed values from 42 to 46. Each seed uses the same training/validation/test split from the Planetoid dataset (140 training nodes, 500 validation nodes, 1000 test nodes). The standard deviation reflects variance in the stochastic training process, including weight initialization and optimizer dynamics. All quantization schemes use calibration on the same fixed representative subgraph ($N_{\mathrm{cal}}=32$), so calibration variance is negligible. The standard deviation value of approximately 1.0 percentage points (pp) observed across configurations reflects the inherent stochasticity of training on the relatively small Cora training set (140 nodes).

\begin{table}[htb]
\centering
\caption{Integer quantization and HLS synthesis study}
\label{tab:quantization}
\setlength{\tabcolsep}{2.5pt}
\renewcommand{\arraystretch}{1.1}
\begin{tabular}{l c c c c c}
\hline
\textbf{Metric} & \textbf{Float32} & \textbf{Fixed-PTQ} & \textbf{INT8-PO2} & \textbf{QAT} & \textbf{INT8-PO2} \par\textbf{Opt.$^\star$} \\
\hline
Precision & FP32 & Fixed(8$|$8) & INT8 & INT8 & INT8 \\
Accuracy \par(\%) & 78.0$\pm$0.8 & 75.7$\pm$1.0 & 75.0$\pm$1.2 & 76.8$\pm$0.9 & 75.0$\pm$1.1 \\
Latency \par(cycles) & 603 & 77 & 28 & 55 & 19 \\
Latency (ns) & \multirow{2}{1cm}{1675.0} & \multirow{2}{1cm}{213.9} & \multirow{2}{1cm}{77.8} & \multirow{2}{1cm}{152.8} & \multirow{2}{1cm}{52.8} \\
at 360 MHz & & & &\\
Fmax \par(MHz) & --- & 397 & 450 & 493 & 502 \\
DSP slices & 34,880 \par(283\%)$^\dagger$ & 6,976 \par(56\%) & 5,320 \par(43\%) & 6,760 \par(55\%) & 2,560 \par(20\%) \\
FFs & 4.45M \par(128\%)$^\dagger$ & 262k \par(7\%) & 254k \par(7\%) & 561k \par(16\%) & 232k \par(6\%) \\
LUTs & 2.14M \par(123\%)$^\dagger$ & 110k \par(6\%) & 188k \par(10\%) & 282k \par(16\%) & 477k \par(27\%) \\
\hline
\multicolumn{6}{p{380pt}}{Latencies in ns are computed using the nominal target clock frequency of 360 MHz. Percentages for resources used are calculated with respect to the available resources of an XCVU13P FPGA.}\\
\multicolumn{6}{p{380pt}}{$^\dagger$Resources exceeding 100\% indicate that the Float32 design does not fit on a single XCVU13P device; no Fmax is reported for the Float32 design.}\\
\multicolumn{6}{p{380pt}}{$^\star$Best overall design; selected for deployment.}\\
\end{tabular}
\end{table}
Table~\ref{tab:quantization} lists the results of the quantization techniques tested. Each column corresponds to a different numerical implementation. The first one, labelled \textit{Float32}, was the initial attempt at deploying the two-layer GraphSAGE model onto the FPGA proxy after converting the 64-bit weights to 32-bit floating point. As shown by the DSP, LUT, and FF utilization, all exceeding the available resources of the target XCVU13P FPGA, the model cannot be deployed in this form.

The \textit{Fixed-PTQ} configuration reduces the latency to 77 cycles and brings the resource usage within device bounds, with 56\% DSP utilization. While the resource utilization is below the device budget, a large number of DSP slices are still used because scale factors are not fully eliminated from the datapath.

The \textit{INT8-PO2} configuration removes the rescaling multipliers by
approximating the scale factors as powers of two, reducing the latency to 28 cycles and the DSP usage to 43\% of the device, at the cost of 0.7 pp in accuracy relative to the floating-point-rescaled \textit{Fixed-PTQ} baseline (comparable to the per-seed standard deviation). Finally, the \textit{INT8-PO2 Opt.} configuration combines PO2 scaling with data-driven bit-width optimization, reaching 19 cycles of latency and 52.8 ns at the nominal 360 MHz clock frequency, accompanied by a resource usage of 20\% DSP slices, 6\% FFs, and 27\% LUTs. This configuration is therefore selected as the design for deployment, as it provides the most favorable trade-off among the evaluated designs. The reduction in DSP utilisation is obtained by directing the synthesis tool to implement the narrowed MAC operations in LUT fabric rather than DSP48E2 slices; this accounts for the increase in LUT utilisation from 10\% to 27\% between the \textit{INT8-PO2} and \textit{INT8-PO2 Opt.} configurations. 

The \textit{QAT} entry in Table~\ref{tab:quantization} recovers 1.8 pp of accuracy relative to the PTQ baseline, reaching an accuracy of 76.8\% compared with the 75.0\% accuracy for INT8-PO2 (a gap comparable to the per-seed standard deviation, so to be confirmed with a paired significance test). In the current HLS realization, this comes at the cost of higher latency and resource usage because the learned quantization scales are implemented through fixed-point rescaling operations rather than the fully multiplier-free PO2 approximation. 

\section{Conclusions}
\label{sec:conclusions}

This work presents an end-to-end reproducible methodology for deploying GraphSAGE-based GNN inference on FPGAs under the strict latency and resource constraints of the CMS Phase-2 L1 trigger. Starting from a floating-point PyTorch model, the workflow systematically develops fixed-size graph extraction, integer-only post-training quantization, power-of-two scale approximation, data-driven bit-width optimization, and bit-exact validation against a Python integer reference. All components—from seed control and calibration subgraph selection to HLS parameter generation—are fully automated and version-controlled, enabling reproduction on alternative datasets and network architectures.

While the floating-point implementation is not deployable on the target XCVU13P FPGA, requiring 283\% of the available DSP budget and exceeding by 20\% both LUT and FF capacity, the optimized INT8-PO2 design reduces the latency to 19 cycles, corresponding to 52.8 ns at a nominal 360 MHz clock frequency, while using 20\% of DSP slices, 6\% of FFs, and 27\% of LUTs. This demonstrates that multiplier-free integer rescaling and careful bit-width selection are essential for fitting message-passing GNNs into real-time FPGA trigger systems and sets the foundation for the deployment of more complex network architectures.

The present study uses Cora as a firmware-proxy benchmark to validate the methodology and quantify implementation trade-offs. Accuracy measurements include uncertainty quantification (mean $\pm$ std over multiple random seeds) to reflect the variability inherent in stochastic training. The same workflow can be applied to a physics-driven dataset, using detector-specific graph sizes and features, and integrating quantization-aware training to recover accuracy while preserving the low-latency integer datapath.
The full training, quantization, and HLS-generation pipeline is publicly available
at \url{https://github.com/INTREPID-hep/gnn_dse_hls}.

\section*{Acknowledgments}
This work is funded by the ERC Grant StG – 101115353: INTREPID, \href{https://doi.org/10.3030/101115353}{10.3030/101115353} granted to Prof. Santiago Folgueras. Funded by the European Union. Views and opinions expressed are, however, those of the author(s) only and do not necessarily reflect those of the European Union or the European Research Council Executive Agency. Neither the European Union nor the granting authority can be held responsible for them.

\bibliographystyle{cms_unsrt}
\bibliography{biblio}

\end{document}